\documentclass[orivec]{llncs}

\title{A Constructor-Based Reachability Logic for Rewrite Theories
}
\titlerunning{A Constructor-Based Reachability Logic for Rewrite Theories}

\author{Stephen Skeirik, Andrei Stefanescu and Jos\'e Meseguer}
\institute{University of Illinois at Urbana-Champaign, USA}

\authorrunning{S. Skeirik, A. Stefanescu, and J. Meseguer}

\usepackage{amsmath,amssymb}
\usepackage{wasysym}
\usepackage{stmaryrd}
\usepackage{cancel}
\usepackage{booktabs}
\usepackage{ebproof}
\usepackage{mathpartir}
\usepackage{tabulary}
\usepackage{soul,color}
\usepackage{hyperref}
\usepackage{comment}

\newcommand{\R}{\ensuremath{\mathcal{R}}}

\newcommand{\sra}{\mbox{{\footnotesize $\rightarrow$}}}

\begin{document}
    \maketitle
	\begin{abstract}
		Reachability logic has been applied to $\mathbb{K}$
		rewrite-rule-based language definitions as a
		\emph{language-generic} logic of programs. 
                To be able to verify not just code but also
                \emph{distributed system designs}, a new \emph{rewrite-theory-generic}
                reachability logic is presented and proved sound for a wide class of
		rewrite theories.
                The logic's automation is increased by means of \emph{constructor-based} semantic
		unification, matching, and satisfiability procedures.  New methods for
		proving invariants of possibly never terminating distributed systems
		are developed, and experiments with a prototype implementation
		illustrating the new proof methods are  presented.

	\end{abstract}

\vspace{-5ex}
	
	\section{Introduction}

\vspace{-.5ex}
	
	The main applications of reachability logic to date
	have been as a \emph{language-generic} logic of programs~\cite{DBLP:conf/rta/StefanescuCMMSR14,DBLP:conf/oopsla/StefanescuPYLR16}.
	In these applications, a $\mathbb{K}$ specification of a language's
	operational semantics by means of rewrite rules is assumed as the language's
	``golden semantic standard,'' and a correct-by-construction
	reachability logic for a language so defined is
	automatically obtained \cite{DBLP:conf/oopsla/StefanescuPYLR16}.  
	This method has been effective in
	proving reachability properties for a wide range of programs.
%
%
	
	Although the foundations of reachability logic are very general
	\cite{DBLP:conf/rta/StefanescuCMMSR14,DBLP:conf/oopsla/StefanescuPYLR16},
        the existing theory does not provide straightforward answers to the
	following questions: (1) Could a reachability logic be
	developed to verify not just conventional programs, but also
	\emph{distributed system designs and algorithms} formalized as
	\emph{rewrite theories} in rewriting logic
	\cite{20-years}?  (2) If so, what would be the most
	natural way to conceive such a
	\emph{rewrite-theory-generic} logic?  
       %
        %
        A satisfactory answer to questions (1)--(2)
	would move the verification game from the level of verifying
	\emph{code} to that of verifying \emph{both code and distributed
		system designs}. Since the cost of design errors can be several orders of
	magnitude higher than that of coding errors, answering
	questions (1) and (2) is of practical software
	engineering interest.
	
	Although a first step towards a reachability logic for rewrite
	theories has been taken in \cite{DBLP:conf/birthday/LucanuRAN15}, as
	explained in Section \ref{REL-CONCL} and below, that first step still
	leaves several important questions open.  The most burning one is how
	to prove \emph{invariants}.  Since
	they  are the most basic safety properties, support for proving
	invariants is a \emph{sine qua non} requirement.  As explained
        below, a serious obstacle is
%
%
what we call the \emph{invariant paradox}:
	we cannot verify in this manner \emph{any} invariants of a
	never-terminating system such as, for example, a mutual exclusion
	protocol.

	A second open question is how to best take advantage of the
	wealth of equational reasoning techniques such as matching,
	unification, and narrowing modulo an equational theory $(\Sigma,E)$,
	and of recent results on decidable satisfiability 
	of quantifier-free formulas in initial algebras, e.g.,
	\cite{var-sat-short}
	to \emph{automate} as much as possible reachability logic deduction.
	In this regard, the very general foundations of reachability logic
	---which assume any $\Sigma$-algebra $\mathcal{A}$ with a first-order-definable
	transition relation--- provide no help at all for automation.
	As shown in this work and its prototype
	implementation, if we assume instead  that the model in question is the
	\emph{initial model} $\mathcal{T}_{\mathcal{R}}$ of a rewrite theory
	$\mathcal{R}$ satisfying reasonable assumptions, 
	large parts of the verification effort can be
	automated.
	
	A third important issue is \emph{simplicity}.  
	Reachability logic has eight inference rules
	\cite{DBLP:conf/rta/StefanescuCMMSR14,DBLP:conf/oopsla/StefanescuPYLR16}.  
	Could a reachability logic
	for rewrite theories be simpler?  This work 
	tackles head on these three open questions to provide
	a general reachability logic and a prototype implementation suitable for reasoning about
	properties of \emph{both} distributed systems and programs based on
	their rewriting logic semantics. 
	
	\vspace{1ex}
	
	\noindent {\bf Rewriting Logic in a Nutshell}. A
	distributed system can be designed and
	modeled as a \emph{rewrite theory} $\mathcal{R}=(\Sigma,E,R)$
	\cite{20-years} in the following way: (i) the distributed
	system's \emph{states} are modeled as elements of the initial algebra
	$T_{\Sigma/E}$ associated to the equational theory $(\Sigma,E)$ with
	function symbols $\Sigma$ and equations $E$; and (ii) the system's
	\emph{concurrent transitions} are modeled by rewrite rules $R$,
	which are applied \emph{modulo} $E$. Let us consider 
	the QLOCK \cite{Futatsugi10} mutual
	exclusion protocol, explained in detail in Section \ref{os-prelims}.  QLOCK
	allows an unbounded number of processes, which can be
	identified by numbers.  Such processes can be
	in one of three states: ``normal'' (doing their own thing),
	``waiting'' for a resource, and ``critical,'' i.e., using the
	resource.  Waiting processes enqueue their identifier
	at the end of a waiting queue and can become
	critical when their name appears at the head of the queue.
	A QLOCK state can  be
	represented as a tuple $\mathit{<n\ |\ w\ |\ c\ |\ q>}$
	where $n$, resp. $w$, resp. $c$,
	denotes the set of identifiers for normal, resp. waiting, resp. critical
	processes, and $q$ is the waiting queue.  QLOCK can be
	modeled as a rewrite theory $\mathcal{R}=(\Sigma,E,R)$,
	where 
%
%
       $E$ includes axioms such
	as associativity-commutativity of multiset union, list
	associativity, and identity
	axioms for $\emptyset$ and $\mathit{nil}$. 
	QLOCK's behavior is
	specified by five rewrite rules $R$. 
	Rule $\mathit{w2c}$ below specifies a waiting process $i$ becoming
	critical
	\[\mathit{w2c}: \mathit{<n\ |\ w\, i\ |\ c\ |\ i;q>} \rightarrow \mathit{<n\ |\ w\ |\
		c\,i \ |\ i;q>.}
	\]
	
	\noindent {\bf Reachability Logic in a Nutshell}.
	A reachability logic formula has the form
	$A \rightarrow^{\circledast} B$, with $A$ and $B$ state
	predicates (see Section \ref{CCP}).  Assume for simplicity
        that  $\mathit{vars}(A) \cap
	\mathit{vars}(B) = \emptyset$.  Such a formula is then  interpreted
	in the initial model  $\mathcal{T}_{\mathcal{R}}$ of a rewrite theory
	$\mathcal{R}=(\Sigma,E,R)$, whose states are $E$-equivalence
	classes $[u]$ of ground $\Sigma$-terms, and
	where a state transition $[u] \rightarrow_{\mathcal{R}} [v]$
	holds iff $\mathcal{R} \vdash u \rightarrow v$
	according to the rewriting logic inference system \cite{20-years}
	(computation = deduction).
	As a first approximation, $A \rightarrow^{\circledast} B$
	is a Hoare logic \emph{partial correctness} assertion of the
	form
%
%
	$\{A\} \mathcal{R} \{B\}$, but with the slight twist
	that $B$ need not hold of a terminating state, but just
	\emph{somewhere along the way}.  To be fully precise,
	$A \rightarrow^{\circledast} B$ holds in $\mathcal{T}_{\mathcal{R}}$
	iff for each state $[u_{0}]$ satisfying $A$ and
	each terminating sequence
	$[u_{0}] \rightarrow_{\mathcal{R}} [u_{1}] \ldots \rightarrow_{\mathcal{R}} [u_{n-1}]
	\rightarrow_{\mathcal{R}} [u_{n}]$ there is a $j$, $0 \leq j \leq n$
	such that $[u_{j}]$ satisfies $B$.  A key question is how to choose
	a good language of state predicates like $A$ and $B$.
	Here is where the potential for increasing the logic's automation
	resides.  We call our proposed logic \emph{constructor-based},
	because our choice is to make $A$ and $B$
	positive (only $\vee$ and $\wedge$) combinations
	of what we call \emph{constructor patterns} of the
	form $u \mid \varphi$, where $u$ is a \emph{constructor}
	term\footnote{That is, a term in a subsignature $\Omega \subseteq \Sigma$ 
		such that each ground $\Sigma$-term is equal modulo $E$ to
		a ground $\Omega$-term.}
	and
	$\varphi$ a quantifier-free (QF) $\Sigma$-formula.
	The state predicate $u \mid \varphi$
	holds for a state $[u']$ iff there is a ground
	substitution $\rho$ such that $[u']=[u \rho]$
	and $E \models  \varphi \rho$.

	\vspace{1ex}
	
	\noindent {\bf The Invariant Paradox}.  How
	can we \emph{prove invariants} in such a reachability logic?  For
	example, mutual exclusion for QLOCK? Paradoxically, we cannot!  This is because
	QLOCK, like many other protocols, \emph{never
		terminates}, that is, has no terminating sequences whatsoever.
	And this has the ludicrous trivial consequence that 
	QLOCK's initial model $\mathcal{T}_{\mathcal{R}}$ vacuously
	satisfies \emph{all} reachability formulas $A
	\rightarrow^{\circledast} B$.  This of course means that it is
	in fact \emph{impossible} to prove any
	invariants using reachability logic in the initial model
	$\mathcal{T}_{\mathcal{R}}$.  But it does \emph{not} mean that it
	is impossible using some other initial model.   In Section \ref{inv-co-inv}
	we give a systematic solution to this paradox
	by means of a \emph{simple theory transformation}
	allowing us to prove any invariant in the original
	initial model $\mathcal{T}_{\mathcal{R}}$ by proving
	an equivalent reachability formula in the initial model of
	the transformed theory.
	
	\vspace{1ex}
	
	\noindent {\bf Our Contributions}.  
	Section \ref{os-prelims} gathers preliminaries.  The main theoretical contributions of a
	\emph{simple} semantics and  inference system for a
	rewrite-theory-generic 
	reachability
	logic with just \emph{two} inference rules
	and its soundness are developed in Sections
	\ref{C-B-REACH} and \ref{sound-inf}.  A systematic methodology
	to prove \emph{invariants} by means of reachability formulas is
	developed in Section \ref{inv-co-inv}.
	The goal of increasing the logic's potential for
	automation by making it constructor-based is advanced in Sections
	\ref{CCP}--\ref{sound-inf}.  A proof of concept of the entire approach is
	given by means of  a Maude-based prototype implementation and a suite of
	experiments verifying various properties of  distributed system designs
	in Section \ref{REFL-IMP}.  Related work and conclusions are
	discussed in Section \ref{REL-CONCL}.
	Proofs can be found in \cite{maude-reach-log}.

\vspace{-2ex}
	
	\section{Many-Sorted Algebra and Rewriting Logic} \label{os-prelims}

\vspace{-.5ex}
	
	We present some preliminaries on many-sorted algebra
	and rewriting logic.
        For a more general treatment using order-sorted algebra see
        \cite{maude-reach-log}. Readers familiar with many-sorted
        logic may go directly to Def. \ref{rewrite-theory-defns}.
       We assume familiarity with the following basic concepts and
       notation that
           are explained in full detail in, e.g., \cite{mg85}:
       (i) {\em many-sorted (MS) signature\/} as a pair $\Sigma = (S,\Sigma)$
		with $S$ a set of \emph{sorts} and $\Sigma$ an $S^{*}
                \times S$-indexed family $\Sigma =
                \{\Sigma_{w,s}\}_{(w,s) \in S^{*} \times S}$ of
                function symbols, where $f \in \Sigma_{s_{1} \ldots
                  s_{n},s}$ is displayed as $f: s_{1} \ldots s_{n}
                \rightarrow s$; (ii) {\em 
			$\Sigma$-algebra\/} $A$ 
                        as a pair $A
                      =(A,\_{_{A}})$ with $A=\{A_{s}\}_{s \in S}$ an
                      $S$-indexed family of sets, and $\_{_{A}}$ a
                      mapping interpreting each $f: s_{1} \ldots s_{n}
                \rightarrow s$ as a function in the set
                $[A_{s_{1}} \times \ldots \times A_{s_{n}}\rightarrow A_s]$. (iii) \emph{$\Sigma$-homomorphism\/} $h: A \rightarrow B$ as
		an $S$-indexed family of functions $h =\{h_{s}: A_{s}
                \rightarrow B_{s}\}_{s \in S}$ preserving the
                operations in $\Sigma$; (iv) the term $\Sigma$-algebra
                $T_{\Sigma}$ and its initiality  in the category ${\bf
                  MSAlg}_{\Sigma}$ of $\Sigma$-algebras when $\Sigma$ is unambiguous.
	
	An $S$-sorted set $X=\{X_{s}\}_{s \in S}$ of \emph{variables},
	satisfies $s \not= s' \Rightarrow X_{s}\cap
	X_{s'}=\emptyset$, and the variables in $X$ are always assumed
	\emph{disjoint} from all constants in $\Sigma$.  The $\Sigma$-\emph{term
		algebra} on variables $X$,
	$T_{\Sigma}(X)$, is the \emph{initial algebra} for the signature
	$\Sigma(X)$ obtained by adding to $\Sigma$ the variables $X$ \emph{as
		extra constants}.  Since a $\Sigma(X)$-algebra is just a pair
	$(A,\alpha)$, with $A$ a $\Sigma$-algebra, and $\alpha$ an
	\emph{interpretation of the constants} in $X$, i.e.,
	an $S$-sorted function $\alpha \in [X \sra A]$, the
	$\Sigma(X)$-initiality of
	$T_{\Sigma}(X)$ means that 
		for each $A \in {\bf MSAlg}_{\Sigma}$ and
		$\alpha \in [X \sra A]$, there exists 
		a unique $\Sigma$-homomorphism, 
		$\_\alpha : T_{\Sigma}(X) \rightarrow A$ extending $\alpha$, i.e.,
		such that for each $s \in S$ and $x \in X_{s}$ we have $x
		\alpha_{s} = \alpha_{s}(x)$.
	 In particular, when $A=T_{\Sigma}(Y)$, an interpretation of
	the constants in $X$, i.e., 
	an $S$-sorted function $\sigma\in[X \sra T_{\Sigma}(Y)]$ is called
	a \emph{substitution}, 
	and its unique homomorphic extension $\_\sigma : T_{\Sigma}(X)
	\rightarrow T_{\Sigma}(Y)$ is also called a substitution.
	Define $\mathit{dom}(\sigma)=\{x \in X \mid x \not= x \sigma\}$,
	and $\mathit{ran}(\sigma)= \bigcup_{x \in \mathit{dom}(\sigma)}
	\mathit{vars}(x \sigma)$. Given variables $Z$, the substitution
	$\sigma|_{Z}$ agrees with $\sigma$ on $Z$ and
	is the identity elsewhere.

       We also assume familiarity with many-sorted first-order logic including: (i) the  first-order language of 
	$\Sigma$-\emph{formulas} for $\Sigma$ a signature (in our case
        $\Sigma$ has only function symbols and the $=$ predicate);
        (ii) given a $\Sigma$-algebra $A$, a formula $\varphi \in \mathit{Form}(\Sigma)$,
	and an assignment $\alpha \in [Y \sra A]$, with
	$Y=\mathit{fvars}(\varphi)$ the free variables of $\varphi$,
	the \emph{satisfaction
		relation} $A,\alpha \models \varphi$ 
       (iii) the notions of a formula $\varphi \in
       \mathit{Form}(\Sigma)$ being \emph{valid}, denoted $A \models
       \varphi$, resp. \emph{satisfiable} in a $\Sigma$-algebra $A$.
      For a subsignature $\Omega \subseteq
	\Sigma$ and $A \in {\bf
		MSAlg}_{\Sigma}$, the \emph{reduct} $A|_{\Omega} \in {\bf
		MSAlg}_{\Omega}$ agrees with
	$A$ in the interpretation of all sorts and operations in $\Omega$ and
	discards everything in $\Sigma \setminus \Omega$.
	If $\varphi \in \mathit{Form}(\Omega)$ we have
	the equivalence $A \models \varphi \; \Leftrightarrow \; A|_{\Omega} \models \varphi$.

	An MS \emph{equational theory}
	is a pair $T=(\Sigma,E)$, with $E$ a set of (possibly conditional) $\Sigma$-equations.
	${\bf MSAlg}_{(\Sigma,E)}$ denotes the full subcategory
	of ${\bf MSAlg}_{\Sigma}$ with objects those $A \in {\bf
		MSAlg}_{\Sigma}$ such that $A \models E$, called the $(\Sigma,E)$-\emph{algebras}.
	${\bf MSAlg}_{(\Sigma,E)}$ has an
	\emph{initial algebra} $T_{\Sigma/E}$ \cite{mg85}.
	The inference system in \cite{mg85} is
	\emph{sound and complete} for MS  equational deduction, i.e., for
	any MS equational theory
	$(\Sigma,E)$, and $\Sigma$-equation $u=v$
	we have an equivalence $E \vdash u=v \; \Leftrightarrow \; E
	\models u=v$. For the sake of simpler inference we assume
        \emph{non-empty sorts}, i.e., $\forall s \in S\; T_{\Sigma},s \not= \emptyset$.
       Deducibility $E \vdash u=v$ is abbreviated 
	as $u =_{E} v$, called $E$-\emph{equality}.
	An $E$-\emph{unifier} of a system of
	$\Sigma$-equations, i.e., a conjunction 
	$\phi=u_{1}=v_{1} \, \wedge \, \ldots \, \wedge \,  u_{n}=v_{n}$ of
	$\Sigma$-equations is a substitution $\sigma$ such that $u_{i} \sigma=_{E} v_{i}\sigma$,
	$1 \leq i \leq n$.  An $E$-\emph{unification algorithm} for
	$(\Sigma,E)$ 
	is an algorithm generating a \emph{complete set} of $E$-unifiers
	$\mathit{Unif}_{E}(\phi)$ for any system of $\Sigma$ equations $\phi$,
	where ``complete'' means that for any $E$-unifier $\sigma$ of $\phi$ there
	is a $\tau \in \mathit{Unif}_{E}(\phi)$ and a substitution
	$\rho$ such that $\sigma =_{E} 
	(\tau \rho)|_{\mathit{dom}(\sigma) \cup \mathit{dom}(\tau)}$,
	where $=_{E}$ here means that for any variable $x$ we have 
	$x\sigma =_{E} x (\tau \rho)|_{\mathit{dom}(\sigma) \cup \mathit{dom}(\tau)}$.
	The algorithm is \emph{finitary} if it  always terminates
	with a \emph{finite set} $\mathit{Unif}_{E}(\phi)$ for any $\phi$.
	
	We recall some basic concepts about \emph{rewriting logic}.  The survey
	in \cite{20-years} gives a fuller  account.  A rewrite theory $\mathcal{R}$ axiomatizes a \emph{distributed
		system}, so that concurrent computation is modeled as concurrent
	rewriting with the rules of $\mathcal{R}$
	\emph{modulo} the equations of $\mathcal{R}$. Recall also the following notation 
	from \cite{dershowitz-jouannaud}: (i) positions in a term viewed as a tree are
	marked by strings $p \in \mathbb{N}^{*}$ specifying
	a path from the root, (ii)  $t|_{p}$ denotes the
	subterm of term $t$ at position $p$, and (iii) $t[u]_{p}$ denotes the result
	of \emph{replacing} subterm $t|_{p}$ at position $p$ by $u$.
	
	\begin{definition} \label{rewrite-theory-defns}
		A \emph{rewrite theory} is a 3-tuple
		$\mathcal{R}=(\Sigma,E \cup B,R)$ 
		%
		%
		with $(\Sigma,E \cup B)$ an MS equational
		theory and $R$ a set of  conditional $\Sigma$-\emph{rewrite rules}
		$l \rightarrow r \; \mathit{if}\; \phi$, with
		$l,r \in T_{\Sigma}(X)_{s}$ for some $s \in S$,
		and $\phi$ a quantifier-free $\Sigma$-formula.
                    %
		%
		%
		We
		further assume that:
		(1)  Each equation $u=v \in B$ is \emph{regular}, i.e., $\mathit{vars}(u) =
			\mathit{vars}(v)$, and \emph{linear}, i.e., there are no repeated variables 
			in either $u$ or $v$. 
			(2) The equations $E$, when oriented as conditional
			rewrite rules $\vec{E}=\{u \rightarrow v \; \mathit{if}\; \psi \mid
			u=v \; \mathit{if}\; \psi \in E\}$,
			are \emph{convergent} modulo $B$,
			i.e., strictly coherent, confluent, and operationally
			terminating as rewrite rules modulo $B$
			\cite{lucas-meseguer-normal-th-JLAMP}.
			(3)  The rules $R$ are \emph{ground coherent} with the equations
			$E$ modulo $B$ \cite{DBLP:journals/jlp/DuranM12}.
			%
			%
			%
\end{definition}
	
        Conditions (1)--(2) ensure that the initial algebra 
	$T_{\Sigma/E \cup B}$ is isomorphic to the \emph{canonical term
		algebra} $C_{\Sigma/E,B}$, whose elements are $B$-equivalence
	classes of $\vec{E},B$-irreducible ground $\Sigma$-terms.
	Define the
	\emph{one-step}
	$R,B$-\emph{rewrite relation} $t \rightarrow_{R,B} t'$ 
	%
	%
	%
	between
	ground terms as follows.
	For $t,t'\in T_{\Sigma,s}$, $s\in S$,
	$t \rightarrow_{R,B} t'$ holds
	iff there is a
	rewrite rule $l\rightarrow r  \; \mathit{if}\; \phi  \in R$, a ground substitution
	$\sigma \in [Y \sra T_{\Sigma}]$ with $Y$ the rule's variables, 
	and a term position $p$ in $t$
	%
	%
	%
	such that $t|_{p}=_{B} l \sigma$,  $t'=t[r \sigma]_{p}$, and 
	$E \cup B \models \phi \sigma$.
	%
	%
	%
	In  the context of (1)--(2), condition (3)  ensures that
	``computing $\vec{E},B$-canonical forms before performing $R,B$-rewriting'' is a
	\emph{complete} strategy.  That is, if $t \rightarrow_{R,B} t'$
	and $u = t!_{E,B}$, i.e., $t \rightarrow^{*}_{\vec{E},B} u$ with $u$ in
	$\vec{E},B$-canonical form (abbreviated in what follows to $u = t!$),
	then there exists a $u'$ such that $u \rightarrow_{R,B} u'$
	and $t'! =_{B} u'!$.  Note that $\mathit{vars}(r) \subseteq
	\mathit{vars}(l)$ \emph{is nowhere assumed} for rules $l\rightarrow r  \;
	\mathit{if}\; \phi  \in R$. This means that $\mathcal{R}$
	can specify an \emph{open system}, in the sense
	of \cite{rw-SMT-JLAMP}, 
	that interacts with an external, non-deterministic environment
        such as, for example, a thermostat.
	
	Conditions (1)--(3) allow a simple
	description of the \emph{initial reachability model}
	$\mathcal{T}_{\mathcal{R}}$ \cite{20-years} of $\mathcal{R}$
	as the
	\emph{canonical reachability model}
	$\mathcal{C}_{\mathcal{R}}$
	whose states belong to the \emph{canonical term algebra} 
	$C_{\Sigma/E,B}$, and 
	the one-step transition relation $[u] \rightarrow_{\mathcal{R}} [v]$
	holds iff $u \rightarrow_{R,B} u'$ and $[u'!] = [v]$.
	Furthermore, if $u \rightarrow_{R,B} u'$ 
	has been performed with a rewrite rule $l\rightarrow r  \;
	\mathit{if}\; \phi  \in R$
	and a ground substitution
	$\sigma \in [Y \sra T_{\Sigma}]$, then, assuming $B$-equality is decidable,
	checking whether condition $E \cup B
	\models \phi \sigma$ holds is \emph{decidable} by reducing the terms in $\phi
	\sigma$ to $\vec{E},B$-canonical form.
	
	\vspace{1ex}
	
	\noindent\textbf{A Running Example.}
	Consider the following rewrite theory 
	$\mathcal{R}=(\Sigma,E \cup B,R)$ modeling a dynamic version of the
	QLOCK mutual exclusion protocol \cite{Futatsugi10}, where
	$(\Sigma,B)$ defines the protocol's states, involving
	natural numbers, lists, and multisets over
	natural numbers.  $\Sigma$  has sorts $S=\{\mathit{Nat},
	\mathit{List},
	\mathit{MSet},\mathit{Conf},\mathit{State},\mathit{Pred}\}$ 
	with subsorts\footnote{As pointed out at the beginning of
          Section \ref{os-prelims}, \cite{maude-reach-log} treats the
          more general \emph{order-sorted} case, where sorts form
          a poset $(S,\leq)$ with $s \leq s'$ interpreted as set
          containment $A_{s} \subseteq A_{s'}$ in a $\Sigma$-algebra $A$.}
		$\mathit{Nat}  < \mathit{List}$ and
	$\mathit{Nat}  < \mathit{MSet}
	$ and operators $F=\{
	\mathit{0} :\ \rightarrow\mathit{Nat},\
	\mathit{s\_}:\mathit{Nat} \rightarrow \mathit{Nat},\
	\emptyset:\ \rightarrow\mathit{MSet},\
	\mathit{nil}:\ \rightarrow\mathit{List},$
	$\mathit{\_\_} : \mathit{MSet\ MSet}\rightarrow\mathit{MSet},\
	\mathit{\_;\_}:\mathit{List\ List}\rightarrow\mathit{List},
	\mathit{dupl}:\mathit{MSet} \rightarrow \mathit{Pred},\mathit{tt}:\
	\rightarrow \mathit{Pred},
	\mathit{<\_>}:\mathit{Conf}\rightarrow\mathit{State},
	\mathit{\_|\_|\_|\_}:\mathit{MSet\ MSet\ MSet\ List}\rightarrow \mathit{Conf}
	\}$,
	where underscores denote operator argument placement.
	The axioms $B$ are the associativity-commutativity
	of the multiset union $\mathit{\_\_}$ 
	with identity $\emptyset$, and the associativity of list concatenation
	$\mathit{\_;\_}$ with identity $\mathit{nil}$.
	The only equation in $E$ is $\mathit{dupl(s\, i \, i)}=\mathit{tt}$.  It
	defines the $\mathit{dupl}$ predicate by  detecting a duplicated
	element $i$ in the multiset $s\, i \, i$ ($s$ could be empty).
	\emph{States} of QLOCK are $B$-equivalence classes
	of ground terms of sort $\mathit{State}$.
	
	QLOCK \cite{Futatsugi10} is a mutual exclusion protocol where the number
	of processes is unbounded.  Furthermore, in the
	\emph{dynamic} version of  QLOCK  presented below, such a number
	can grow or shrink.
	Each process is identified by a number.  The system
	configuration has three sets of processes (normal, waiting, and critical)
	plus a waiting queue.  To ensure mutual exclusion,
	a normal process must first register its name at the end of the waiting queue.
	When its name appears at the front of the queue, it is allowed to enter the critical section.
	The first three rewrite rules in $R$ below specify how a
	\emph{normal} process $i$ first transitions to a \emph{waiting} process,
	then to a \emph{critical} process, and back to normal.  The last two
	rules in $R$ specify how a
	process can dynamically join or exit the system.
	
	\begin{center}
		$\mathit{n2w} :
		\mathit{<n\ i\           |\ w\hspace*{11pt} |\ c\hspace*{11pt} |\ q\hspace*{13.5pt} >}\ \rightarrow\
		\mathit{<n\hspace*{10.5pt} |\ w\ i\           |\ c\hspace*{11pt} |\ q\ ;\, i          >}\hspace*{21pt}$\\
		$\mathit{w2c}\hspace*{1pt} :
		\mathit{<n\hspace*{11pt} |\ w\ i\           |\ c\hspace*{11pt} |\ i\ ;\, q          >}\ \rightarrow\
		\mathit{<n\hspace*{11pt} |\ w\hspace*{11pt} |\ c\ i\           |\ i\ ;\, q          >}\hspace*{21pt}$\\
		$\mathit{c2n}\hspace*{1.5pt} :
		\mathit{<n\hspace*{11pt} |\ w\hspace*{11pt} |\ c\ i\           |\ i\ ;\, q          >}\ \rightarrow\
		\mathit{<n\ i\           |\ w\hspace*{11pt} |\ c\hspace*{11pt} |\ q\hspace*{13.5pt} >}\hspace*{21pt}$
		$\mathit{join}\hspace*{0.5pt} :
		\mathit{<n\hspace*{11pt} |\ w\hspace*{11pt} |\ c\hspace*{11pt} |\ q\hspace*{13.5pt} >}\ \rightarrow\
		\mathit{<n\ i\           |\ w\hspace*{11pt} |\ c\hspace*{11pt} |\ q\hspace*{13.5pt} >}\ \mathit{if}\ \phi$
		$\mathit{exit}\hspace*{1.5pt} :
		\mathit{<n\ i\           |\ w\hspace*{11pt} |\ c\hspace*{11pt} |\ q\hspace*{13.5pt} >}\ \rightarrow\
		\mathit{<n\hspace*{11pt} |\ w\hspace*{11pt}   |\ c\hspace*{11pt} |\ q\hspace*{13.5pt} >}\hspace*{21pt}$\\
	\end{center}
	
	\noindent where $\phi \equiv \mathit{dupl}(n \, i \, w\ c)\neq\mathit{tt}$,
	$\mathit{i}$ is a number,
	$\mathit{n}$, $\mathit{w}$, and $\mathit{c}$ are, respectively, normal, waiting, and critical
	process identifier sets, and $\mathit{q}$ is a queue of
	process identifiers. 
	It is easy to check that 
	$\mathcal{R}=(\Sigma,E \cup B,R)$ satisfies 
	requirements (1)--(3).  Note that $\mathit{join}$ makes
	QLOCK an \emph{open} system in the sense explained above.
	
\vspace{-2ex}
	
	\section{Constrained Constructor Pattern Predicates} \label{CCP}
	
\vspace{-.5ex}

	Given an MS equational theory $(\Sigma,E \cup B)$,
	the \emph{atomic state predicates} appearing in the constructor-based reachability
	logic formulas of Section \ref{C-B-REACH}
	will be pairs $u \mid \varphi$, called \emph{constrained constructor patterns},
	with $u$ a term in a subsignature $\Omega
	\subseteq \Sigma$
	of constructors, and $\varphi$ a quantifier-free $\Sigma$-formula.
	Intuitively, $u \mid \varphi$ is a pattern describing the set of 
	states that are $E_{\Omega} \cup
	B_{\Omega}$-equal to ground terms
	of the form $u \rho$ for $\rho$ a ground constructor substitution
	such that $E \cup B \models \varphi \rho$.  Therefore, $u \mid \varphi$
	can be used as a \emph{symbolic description} of a, typically
	infinite, \emph{set of states} in the canonical reachability model $\mathcal{C}_{\mathcal{R}}$
	of a rewrite theory $\mathcal{R}$.  
	
	Often, the signature $\Sigma$ on which $T_{\Sigma/E \cup B}$ is defined
	has a natural decomposition as a
	disjoint union $\Sigma = \Omega \uplus \Delta$, where
	the elements of the canonical term algebra $C_{\Sigma/E,B}$
	are $\Omega$-terms,
	whereas the function symbols $f \in  \Delta$ are viewed as
	\emph{defined functions} which are
	\emph{evaluated away} by $\vec{E},B$-simplification.
	$\Omega$ (with same poset of sorts as $\Sigma$)
	is then called a \emph{constructor subsignature} of $\Sigma$.
	
	A \emph{decomposition} of 
	a MS equational theory $(\Sigma,E \cup B)$ is a triple 
	$(\Sigma,B,\vec{E})$ such that the rules $\vec{E}$ are convergent modulo $B$.
	$(\Sigma,B,\vec{E})$ is called
	\emph{sufficiently complete} with
	respect to the \emph{constructor subsignature} $\Omega$
	iff  for each $t \in T_{\Sigma}$ we
	have: (i) $t!_{\vec{E},B} \in T_{\Omega}$, and (ii) if $u \in T_{\Omega}$
	and $u =_{B} v$, then $v \in T_{\Omega}$.  This
	ensures that for each $[u]_{B} \in C_{\Sigma/E,B}$
	we have $[u]_{B} \subseteq T_{\Omega}$.
	Sufficient completeness is closely
	related to the notion of a \emph{protecting}
	inclusion of decompositions.
	
	\begin{definition}  \label{protecting}
		Let $(\Sigma_{0},E_{0} \cup B_{0}) \subseteq (\Sigma,E \cup B)$ 
		be a theory inclusion such that $(\Sigma_{0},B_{0},\vec{E_{0}})$
		and $(\Sigma,B,\vec{E})$ are respective decompositions of 
		$(\Sigma_{0},E_{0} \cup B_{0})$ and $(\Sigma,E \cup B)$.
		We then say that the decomposition
		$(\Sigma,B,\vec{E})$ 
		\emph{protects} $(\Sigma_{0},B_{0},\vec{E_{0}})$
		iff (i) for all  $t,t' \in T_{\Sigma_{0}}(X)$ we have:
		(i) $t=_{B_0} t' \Leftrightarrow t=_{B} t'$, (ii) $t=t!_{\vec{E_0},B_{0}}
		\Leftrightarrow t=t!_{\vec{E},B}$, and (iii)
		$C_{\Sigma_{0}/E_{0},B_{0}}=C_{\Sigma/E,B}|_{\Sigma_{0}}$.
		
		$(\Omega,B_{\Omega},\vec{E_{\Omega}})$ is a
		\emph{constructor decomposition} of $(\Sigma,B,\vec{E})$ iff
		(i) $(\Sigma,B,\vec{E})$ protects $(\Omega,B_{\Omega},\vec{E_{\Omega}})$,
		and (ii) $(\Sigma,B,\vec{E})$ is sufficiently complete with respect to
		the constructor subsignature $\Omega$.
		Furthermore, $\Omega$ is called a subsignature of \emph{free
			constructors modulo} $B_{\Omega}$ iff $E_{\Omega} = \emptyset$, so
		that $C_{\Omega/E_{\Omega},B_{\Omega}}=T_{\Omega/,B_{\Omega}}$.
	\end{definition}

\vspace{-.5ex}
	
	We are now ready to define constrained constructor pattern
        predicates. 

\vspace{-.5ex}
	
	\begin{definition} \label{ctor-pat-pred-defn}
		Let $(\Omega,B_{\Omega},\vec{E_{\Omega}})$ be a
		\emph{constructor decomposition} of $(\Sigma,B,\vec{E})$.  A
		\emph{constrained constructor pattern} is an expression
		$u \mid \varphi$ with $u \in T_{\Omega}(X)$ and $\varphi$ a QF
		$\Sigma$-formula.  The set $\mathit{PatPred}(\Omega,\Sigma)$
		of \emph{constrained constructor pattern \text{predicates}} contains
		$\bot$ and the set of constrained constructor patterns, and
		is closed
		under disjunction ($\vee\!$) and conjunction ($\wedge\!$). 
		Capital letters $A,B,\ldots,P,Q,\ldots$ range over $\mathit{PatPred}(\Omega,\Sigma)$.
		The
		\emph{semantics} of a constrained constructor pattern predicate $A$
		is a subset $\llbracket A \rrbracket \subseteq C_{\Sigma/E,B}$
		defined inductively as follows:
		\begin{enumerate}
			\item $\llbracket \bot \rrbracket = \emptyset$
			\item $\llbracket u \mid \varphi \rrbracket = \{[(u \rho)!]_{B_{\Omega}} \in
			C_{\Sigma/E,B} \mid \rho \in  [X \sra T_{\Omega}] \, \wedge \, E \cup B \models \varphi \rho\}$.
			\item $\llbracket A \vee B \rrbracket$ = $\llbracket A 
			\rrbracket \cup \llbracket B \rrbracket$
			\item $\llbracket A \wedge B \rrbracket$ = $\llbracket A
			\rrbracket \cap \llbracket B \rrbracket$.
		\end{enumerate}
	\end{definition}

\vspace{-.5ex}
	
	Note that for any constructor pattern predicate $A$, if $\sigma$ is a
	(sort-preserving) bijective renaming of variables we always have
	$\llbracket A \rrbracket = \llbracket A \sigma \rrbracket$.
	Given constructor patterns $ u \mid \varphi $ and $v \mid \psi$
	with $\mathit{vars}(u \mid \varphi) \cap \mathit{vars}(v \mid \psi) = \emptyset$,
	we say that $ u \mid \varphi $ \emph{subsumes}
	$v \mid \psi$ iff there is a substitution $\alpha$ such that:
	(i) $v =_{E_{\Omega} \cup B_{\Omega}} u \alpha$, and (ii)
	$\mathcal{T}_{E \cup B} \models \psi \Rightarrow (\varphi \alpha)$.
	It then follows easily from the above definition of $\llbracket u \mid
	\varphi \rrbracket$ that if $ u \mid \varphi $ subsumes
	$v \mid \psi$, then $\llbracket v \mid
	\psi \rrbracket \subseteq \llbracket u \mid
	\varphi \rrbracket$.  Likewise, 
	$\bigvee_{i \in I} u_i \mid \varphi_i $ \emph{subsumes}
	$v \mid \psi$ iff there is a $k \in I$ such that $u_k \mid \varphi_k$
	subsumes $v \mid \psi$.
	
	\vspace{1ex}
	
	\noindent\textbf{Pattern Predicate Example.}
	Letting $\mathit{n}$, $\mathit{w}$,
	$\mathit{c}$ be multisets of process identifiers and $\mathit{q}$ be an
	associative list of process identifiers, recall that QLOCK states have the  form
	$\mathit{<n\ |\ w\ |\ c\ |\ q>}$.  From the five rewrite rules
	defining QLOCK, it is easy to prove that if
	$\mathit{<n\ |\ w\ |\ c\ |\ q>} \rightarrow^* \mathit{<n'\ |\ w'\ |\
		c'\ |\ q'>}$
	and $n \, w \, c$ is a set (has no repeated elements), then
	$n' \, w' \, c'$ is also a set.  Of course, it seems very reasonable
	to assume that these process identifier multisets are, in fact, sets,
	since otherwise we could, for example, have a process $i$ which is
	\emph{both} waiting and critical at the \emph{same} time.
	We can rule out such ambiguous states by means
	of the pattern predicate
	$\mathit{<n\ |\ w\ |\ c\ |\ q>}\ |\
	\mathit{dupl}(\mathit{n\ w\ c})\neq \mathit{tt}$.
	
	\vspace{1ex}
	
	If $E_{\Omega} \cup B_{\Omega}$ has a finitary
	unification algorithm, any constrained constructor
	pattern predicate $A$ is semantically equivalent 
	to a finite disjunction $\bigvee_{i} u_{i} \mid \varphi_{i}$ of constrained
	constructor patterns.  This is
	because: (i) by (3)--(4) in Def. \ref{ctor-pat-pred-defn} we may assume $A$ in disjunctive
	normal form; and (ii) it is easy to check that
	$\llbracket (u \mid \varphi) \wedge (v \mid \phi) \rrbracket$
	= $\bigcup_{\alpha \in \mathit{Unif}_{E_{\Omega} \cup B_{\Omega}}(u,v)} \llbracket u \alpha
	\mid (\varphi \wedge \phi) \alpha \rrbracket$,
	were we assume that
	$\mathit{vars}(u \mid \varphi) \cap \mathit{vars}(v \mid \psi) = \emptyset$,
	and that all variables in $\mathit{ran}(\alpha)$ are
        \emph{fresh}.  Pattern intersection
        can also be defined when $u \mid \varphi$ and  $v \mid \phi$ share \emph{parameters} 
        $Y=\mathit{vars}(u \mid \varphi) \cap \mathit{vars}(v \mid \phi)
	=\mathit{vars}(u) \cap \mathit{vars}(v)$.  \cite{maude-reach-log}
        defines in detail the notions of
        \emph{parametric intersection} $\llbracket u \mid \varphi
	\rrbracket \cap_{Y} \llbracket  v \mid \phi   \rrbracket$ and
        of \emph{parametric subsumption} $v \mid \phi \subseteq_{Y} u
        \mid \varphi$ of patterns. These notions are very useful 
	 to reason about \emph{parameterized}
	invariants and co-invariants (see Section \ref{inv-co-inv} and
        \cite{maude-reach-log}).
	
\vspace{-2ex}
	
	\section{Constructor-Based Reachability Logic} \label{C-B-REACH}

\vspace{-.5ex}
	
	The constructor-based reachability logic we define is a
	logic to reason about reachability properties of the canonical
	reachability model  $\mathcal{C}_{\mathcal{R}}$
	of a topmost rewrite theory $\mathcal{R}$ where ``topmost'' 
	intuitively means all rewrites must occur at the top of the
        term.\footnote{Topmost theories have reachability completeness
          for narrowing \cite{narrowing-hosc}.  Our inference system uses narrowing to symbolically compute successor states in $\mathcal{C}_\R$.}  Many rewrite theories
	of interest, including those specifying distributed object-oriented systems or the semantics of (possibly concurrent) programming
	languages, can be easily made topmost by a theory transformation (see, e.g., \cite{narrowing-hosc}).
	Formally, we require $\mathcal{R}=(\Sigma,E \cup B,R)$, besides satisfying the requirements in
	Definition
	\ref{rewrite-theory-defns}, also satisfies:
	
	\begin{enumerate}
		\item $(\Sigma,E \cup B)$ has a
		sort $\mathit{State}$, 
		a decomposition $(\Sigma,B,\vec{E})$,
		and a constructor decomposition
		$(\Omega,B_{\Omega},\vec{E_{\Omega}})$ where:
		(i) $\forall u \in T_{\Omega}(X)_{\mathit{State}}$, $\!\mathit{vars}(u)=\mathit{vars}(u!)$;
		%
		%
		%
		(ii) $B_{\Omega}$ are linear and regular with
		a finitary $E_{\Omega} \cup B_{\Omega}$-unification
		algorithm. 

		\item Rules in $R$ have the form 
		$l\rightarrow r  \;
		\mathit{if}\; \varphi$ with $l \in T_{\Omega}(X)$.  Furthermore, they
		are \emph{topmost} in the sense that: (i) for all such rules,
		$l$ and $r$ have sort $\mathit{State}$,
		and (ii) for any $u \in T_{\Omega}(X)_{\mathit{State}}$
		and any non-empty position $p$ in $u$, 
		$u|_{p} \not\in T_{\Omega}(X)_{\mathit{State}}$.
	\end{enumerate}

	Requirements
	(1)--(2) ensure that in the canonical reachability model
	$\mathcal{C}_{\mathcal{R}}$ if $[u] \rightarrow_{\mathcal{R}} [v]$
	holds, then the $R,B$-rewrite $u \rightarrow_{R,B} u'$ such
	that $[u'!] = [v]$  \emph{happens at the top} of
	$u$, i.e., uses a rewrite rule
	$l\rightarrow r \; \mathit{if}\; \varphi \in R$ and a ground substitution
	$\sigma \in [Y \sra T_{\Omega}]$, with $Y$ the rule's variables, such that
	$u =_{B_{\Omega}} l \sigma$ and $u'=r \sigma$.
	
	We are now ready to define the formulas of our constructor-based 
	reachability logic for $\mathcal{R}$ satisfying above requirements
	(1)--(2). Let $\mathit{PatPred}(\Omega,\Sigma)_{\mathit{State}}$
	denote the subset of  $\mathit{PatPred}(\Omega,\Sigma)$
	determined by those pattern predicates $A$ such that, for
	all atomic constrained constructor predicates
	$u \mid \varphi$ appearing in $A$, $u$ has sort $\mathit{State}$.
	Reachability logic formulas then have the form:
	$A \rightarrow^{\circledast} B$, with
	$A,B \in \mathit{PatPred}(\Omega,\Sigma)_{\mathit{State}}$.
	The \emph{parameters} $Y$ of 
	$A \rightarrow^{\circledast} B$ are the variables in the set
	$Y=\mathit{vars}(A) \cap \mathit{vars}(B)$, and $A
	\rightarrow^{\circledast} B$ is called \emph{unparameterized}
	iff $Y=\emptyset$.

	The reachability logic in
	\cite{DBLP:conf/rta/StefanescuCMMSR14,DBLP:conf/oopsla/StefanescuPYLR16} 
	is based on
	\emph{terminating} sequences of state transitions;
	when there are no terminating states, \emph{all} 
	reachability formulas are \emph{vacuously true}.
	Our purpose is to extend the logic in order to
	verify properties of general distributed systems specified as rewrite
	theories $\mathcal{R}$ which \emph{may never terminate}.
	For this, as explained in Section \ref{inv-co-inv}, 
	we
	\emph{generalize} the  \emph{all-paths} satisfaction relation
        in \cite{DBLP:conf/oopsla/StefanescuPYLR16}, which for a
        theory $\mathcal{R}$ we denote by
	$\mathcal{R} \models^{\forall} A \rightarrow^{\circledast}
	B$, to a \emph{relativized} satisfaction relation
	$\mathcal{R} \models_{T}^{\forall} A \rightarrow^{\circledast} B$,
	where $T$ is a constrained pattern predicate such that $\llbracket T
	\rrbracket$ is a set of terminating states.
	That is, let $\mathit{Term}_{\mathcal{R}} = \{ [u] \in
	\mathcal{C}_{\mathcal{R},\mathit{State}} \mid (\not\exists[v])\; [u] \rightarrow_{\mathcal{R}}
	[v]\}$.  We then require $\llbracket T \rrbracket \subseteq \mathit{Term}_{\mathcal{R}}$.
	The standard relation $\mathcal{R} \models^{\forall} A \rightarrow^{\circledast}
	B$ is then recovered as the special case where $\llbracket T \rrbracket = \mathit{Term}_{\mathcal{R}}$.
	Call $[u] \rightarrow^{*}_{\mathcal{R}}[v]$
	a $T$-\emph{terminating sequence}
	iff $[v] \in \llbracket T \rrbracket$.

	\begin{definition} \label{all-paths-semantics}
		Given $T$ with $\llbracket T \rrbracket \subseteq \mathit{Term}_{\mathcal{R}}$,
		the \emph{all-paths} satisfaction relation
		$\mathcal{R} \models_{T}^{\forall} u \mid \varphi
		\rightarrow^{\circledast} \bigvee_{j \in J} v_{j} \mid \phi_{j}$
		asserts the satisfaction of the formula
		$u \mid \varphi
		\rightarrow^{\circledast} \bigvee_{j \in J} v_{j} \mid \phi_{j}$
		in the canonical
		reachability model $\mathcal{C}_{\mathcal{R}}$ of a rewrite theory
		$\mathcal{R}$ satisfying topmost  requirements (1)--(2).  It is defined as follows:
		
		For $u \mid \varphi
		\rightarrow^{\circledast} \bigvee_{j \in J} v_{j} \mid \phi_{j}$ 
		unparameterized,
		$\mathcal{R} \models_{T}^{\forall} u \mid \varphi
		\rightarrow^{\circledast} \bigvee_{j \in J} v_{j} \mid \phi_{j}$ holds iff
		for each
		$T$-terminating sequence
		$[u_{0}] \rightarrow_{\mathcal{R}}[u_{1}] \ldots
		[u_{n-1}]\rightarrow_{\mathcal{R}}[u_{n}]$
		with $[u_{0}]  \in \llbracket  u \mid \varphi \rrbracket$ 
		there exist
		$k$, $0 \leq k \leq n$ and $j \in J$ such that
		$[u_{k}] \in \llbracket  v_{j} \mid \phi_{j} \rrbracket$.
		For $u \mid \varphi
		\rightarrow^{\circledast} \bigvee_{j \in J} v_{j} \mid \phi_{j}$ 
		with parameters $Y$, 
		$\mathcal{R} \models_{T}^{\forall} u \mid \varphi
		\rightarrow^{\circledast} \bigvee_{j \in J} v_{j} \mid \phi_{j}$
		holds if
		$\mathcal{R} \models_{T}^{\forall} (u \mid \varphi)\rho
		\rightarrow^{\circledast} (\bigvee_{j \in J} v_{j} \mid \phi_{j})\rho$ holds
		for each $\rho \in [Y \sra T_{\Omega}]$.
		
		Since a constrained pattern predicate is 
		equivalent to a disjunction of atomic ones, we can define 
		satisfaction on  general reachability logic formulas
		as follows:  $\mathcal{R} \models_{T}^{\forall} \bigvee_{1 \leq i \leq
			n} u_i \mid \varphi_i
		\rightarrow^{\circledast} A$ iff
		$\bigwedge_{1 \leq i \leq n} \mathcal{R} \models_{T}^{\forall} u_i \mid \varphi_i
		\rightarrow^{\circledast} A$,
		assuming same parameters $Y_{i}=\mathit{vars}(u_i \mid \varphi_i) \cap \mathit{vars}(A)$,
		i.e., $Y_{i}=Y_{i'}$ for $1 \leq i < i' \leq n$.
	\end{definition}
	
	$\mathcal{R} \models_{T}^{\forall} A \rightarrow^{\circledast} B$
	is a \emph{partial correctness assertion}: If state $[u]$
	satisfies ``precondition''  $A$, then ``postcondition'' $B$ is
	satisfied \emph{somewhere} along \emph{each} $T$-terminating sequences from
	$[u]$, generalizing a Hoare formula $\{A\} \mathcal{R} \{B\}$
        \cite{maude-reach-log}.
	
	Recall that rewrite rules 
	$l\rightarrow r  \;
	\mathit{if}\; \phi$ are assumed to have $l \in T_{\Omega}(X)$.
	For symbolic reasoning purposes it will be very useful to 
	also require that $r \in T_{\Omega}(X)$.
	This can be achieved
	by a theory transformation
	$\mathcal{R} \mapsto \hat{\mathcal{R}}$.
    Stated formally, if $\mathcal{R}=(\Sigma,E \cup B,R)$, then
    $\hat{\mathcal{R}}=(\Sigma,E \cup B,\hat{R})$,
    where the rules $\hat{R}$ are obtained from the rules $R$
    by transforming each $l\rightarrow r  \;
    \mathit{if}\; \phi$ in $R$ into the rule
    $l\rightarrow r'  \;
    \mathit{if}\; \phi \wedge \hat{\theta}$, where:
    (i) $r'$ is the $\Omega$-\emph{abstraction} of $r$ obtained by
    replacing each length-minimal position $p$ of $r$ such that 
    $t|_{p} \not\in T_{\Omega}(X)$ by a fresh variable $x_{p}$ 
    whose sort is the least sort of $t|_{p}$, (ii) $\hat{\theta}
    = \bigwedge_{p\in P} x_{p} = t_{p}$, where $P$ is the set of
    all length-minimal positions in $r$ such that $t|_{p} \not\in
    T_{\Omega}(X)$.    
        
	The key semantic property about this transformation  is:
	
	\begin{lemma} \label{R-hat-transf}
		The canonical reachability models $\mathcal{C}_{\mathcal{R}}$ and
		$\mathcal{C}_{\hat{\mathcal{R}}}$ are \emph{identical}.
	\end{lemma}

\vspace{-3ex}
	
	\subsection{Invariants, Co-Invariants, and Never-Terminating
		Systems} \label{inv-co-inv}

\vspace{-.5ex}
	
	The notion of an \emph{invariant} applies to any transition
	system $\mathcal{S} = (S,\rightarrow_{\mathcal{S}})$ with \emph{states}  $S$ 
	and \emph{transition relation} $\rightarrow_{\mathcal{S}} \subseteq S \times S$.
        The set $\mathit{Reach}(S_{0})$ of states
	\emph{reachable} from $S_{0} \subseteq S$ is defined as
	$\mathit{Reach}(S_{0}) = \{s \in S \mid (\exists s_{0} \in S_{0})\;
	s_{0} \rightarrow^{\ast}_{\mathcal{S}} s \}$, where
	$\rightarrow^{\ast}_{\mathcal{S}}$ denotes the reflexive-transitive closure of
	$\rightarrow_{\mathcal{S}}$.  An invariant 
	about $\mathcal{S}$ with initial states $S_{0}$ can be specified
	in two ways: (i) by a ``good'' property $P \subseteq S$, the \emph{invariant}, that
	\emph{always holds} from $S_{0}$, i.e., such that $\mathit{Reach}(S_{0})
	\subseteq P$, or (ii) as a ``bad'' property $Q \subseteq S$, 
	the \emph{co-invariant},  that \emph{never holds} from $S_{0}$, i.e., such that $\mathit{Reach}(S_{0})
	\cap Q = \emptyset$.  Obviously, $P$ is an invariant iff $S \setminus P$ is a co-invariant.
%

%
%
	  Suppose we
	have specified a distributed system by
	a topmost rewrite theory $\mathcal{R}$, and 
	constrained pattern predicates $S_{0}$ and $P$, and we want to prove that $\llbracket P \rrbracket$
       is an \emph{invariant}  of
	the system
	$(\mathcal{C}_{\mathcal{R},\mathit{State}},\rightarrow_{\mathcal{R}})$
	from $\llbracket S_{0} \rrbracket$.  Can we specify such invariant or
	co-invariant by means of \emph{reachability formulas} and
	use the inference system of Section \ref{sound-inf} to try to prove
	such formulas?
	
	The answer to the above question is not obvious.  Suppose 
	$\mathcal{R}$ specifies a \emph{never-terminating system}, i.e., $\mathit{Term}_{\mathcal{R}}= \emptyset$.  
        For example, QLOCK and other mutual exclusion protocols are
	never-terminating.  Then, no
	reachability formula can characterize and
	invariant  holding 
	by means of the satisfaction relation $\mathcal{R} \models_{T}^{\forall} A \rightarrow^{\circledast} B$.
	The reason for this impossibility is that, since
	$\mathit{Term}_{\mathcal{R}}= \emptyset$,
	$\mathcal{R} \models^{\forall}_{T} A \rightarrow^{\circledast} B$
	holds vacuously for \emph{all} reachability formulas $A
	\rightarrow^{\circledast} B$. 
	
	Is then reachability logic useless
	to prove invariants?  Definitely \emph{not}. 
%
%
         We need
	to first perform a simple \emph{theory transformation}.
	Call an invariant 
	\emph{specifiable by constrained pattern predicates}
	$S_{0}$ and $P$ if $\llbracket P \rrbracket$ is an \emph{invariant}  of
	$(\mathcal{C}_{\mathcal{R},\mathit{State}},\rightarrow_{\mathcal{R}})$
	from $\llbracket S_{0} \rrbracket$.
	%
	%
	%
	To ease the exposition, we explain the transformation for the case
	where $\Omega$ has a single state constructor operator, say,
	$\langle\_,\ldots,\_\rangle : s_{1},\ldots,s_{n} \rightarrow
	\mathit{State}$. The extension to several such operators is
	straightforward. 
	The theory transformation 
	is of the form $\mathcal{R} \mapsto \mathcal{R}_{\mathit{stop}}$,
	where $\mathcal{R}_{\mathit{stop}}$ is obtained from $\mathcal{R}$ by
	just adding: (1) a new state constructor operator
	$[\_,\ldots,\_] : s_{1},\ldots,s_{n}\rightarrow
	\mathit{State}$ to $\Omega$, and (2) a new rewrite
	rule $\mathit{stop} : \langle x_{1}\!\! : \!\!s_{1},\ldots, x_{n}\!\!
	: \!\!s_{n} \rangle \rightarrow [x_{1}\!\! : \!\!s_{1},\ldots,
	x_{n}\!\! : \!\!s_{n}]$ to $R$.  Also, let $[\,]$ denote the
	pattern predicate $ [x_{1}\!\! : \!\!s_{1},\ldots,x_{n}\!\! :
	\!\!s_{n}] \mid \top$.  Likewise, for any atomic 
	constrained pattern predicate 
	$B = \langle u_{1},\ldots, u_{n} \rangle \mid \varphi$
	we define the pattern predicate 
	$[B] = [u_{1},\ldots, u_{n}] \mid \varphi$
	and extend this notation to any union $Q$ of atomic predicates.
	Since $\langle\_,\ldots,\_\rangle : s_{1},\ldots,s_{n}\rightarrow
	\mathit{State}$ is the only state constructor, we can assume without
	loss of generality that any atomic constrained pattern predicate in $\mathcal{R}$ is semantically
	equivalent to one of the form $\langle u_{1},\ldots, u_{n} \rangle \mid \varphi$.
	Likewise, any pattern predicate will be semantically
	equivalent to a union of atomic predicates of such form, called in
	\emph{standard form}.
	%
	%

\vspace{-.5ex}
	
	\begin{theorem} \label{invariant-theorem}
		For $S_{0},P \in \mathit{PatPred}(\Omega,\Sigma)$
		constrained  pattern predicates 
		in standard form with $\mathit{vars}(S_{0}) \cap
		\mathit{vars}(P)= \emptyset$,
		%
		%
		%
		$\llbracket P \rrbracket$ is an invariant  of 
		$(\mathcal{C}_{\mathcal{R},\mathit{State}},\rightarrow_{\mathcal{R}})$
		from $\llbracket S_{0} \rrbracket$ iff 
		$\mathcal{R}_{\mathit{stop}} \models_{[\,]}^{\forall} S_{0}
		\rightarrow^{\circledast} [P]$.
		%
		%
	\end{theorem}
	
	The notion of a \emph{parametric invariant} can be reduced to the
	unparameterized one: 
	if  $Y=\mathit{vars}(S_{0}) \cap
	\mathit{vars}(P)$, then 
	$\llbracket P \rrbracket$ is an \emph{invariant}  of 
	$(\mathcal{C}_{\mathcal{R},\mathit{State}},\rightarrow_{\mathcal{R}})$
	from $\llbracket S_{0} \rrbracket$ \emph{with parameters} $Y$
	iff $\mathcal{R}_{\mathit{stop}} \models_{[\,]}^{\forall} S_{0}
	\rightarrow^{\circledast} [P]$. That is, iff
	$\llbracket P\rho \rrbracket$ is an (unparameterized) invariant  of 
	$(\mathcal{C}_{\mathcal{R},\mathit{State}},\rightarrow_{\mathcal{R}})$
	from $\llbracket S_{0} \rho\rrbracket$ for each $\rho \in [Y \sra
	T_{\Omega}]$.  In this way, Theorem \ref{invariant-theorem} extends
	to parametric invariants.
	
	\vspace{1ex}

	\noindent\textbf{Specifying Invariants for QLOCK.}  Consider the
	QLOCK specification from Sections \ref{os-prelims} and \ref{CCP}. 
	QLOCK nonterminating is \emph{never}
	terminating. 
%
       However, we can apply
	the theory transformation in Theorem \ref{invariant-theorem} by
	adding an operator $\mathit{[\_]}:\mathit{Conf} \rightarrow \mathit{State}$
	and a rule $\mathit{stop}: \,\mathit{<t>}\rightarrow\mathit{[t]}$ for $t\!\!:\!\!\mathit{Conf}$.
	Define the set of initial states by the pattern predicate
	$S_0=\mathit{<n'\ |\ \emptyset\ |\ \emptyset\ |\ nil>} \mid \mathit{dupl}(n') \not= \mathit{tt}$.
	Since QLOCK states have the form
	$\mathit{<n\ |\ w\ |\ c\ |\ q>}$, mutual exclusion means
	$|c|\leq 1$, which is expressible by the pattern predicate
	$\mathit{<n\ |\ w\ |\ i\ |\ i\ ;\ q>} \vee \mathit{<n\ |\ w\ |\ \emptyset\ |\ q>}$.
	But we need also to ensure our multisets are actually \emph{sets}.
	Thus, the pattern predicate
	$P=\big(\mathit{<n\ |\ w\ |\ i\ |\ i\ ;\ q>}\ \! |\ \! \mathit{dupl}(\mathit{n\ w\ i})\neq \mathit{tt}\big)\! \vee\!
	\big(\mathit{<n\ |\ w\ |\ \emptyset\ |\ q>}\; \! |\; \!
	\mathit{dupl}(\mathit{n\ w})\neq \mathit{tt}\big)$
	specifies mutual exclusion.
	By Theorem \ref{invariant-theorem}, QLOCK ensures mutual exclusion
	from $\llbracket S_{0} \rrbracket$ iff $\mathcal{R}_{\mathit{stop}} \models_{[\,]}^{\forall} S_{0}
	\rightarrow^{\circledast} [P]$.
	
	\vspace{1ex}
	
	The following easy corollary
	can be very helpful in proving invariants.  It can, for example,
	be applied to prove the mutual exclusion of QLOCK.
	
	\begin{corollary} \label{invariant-coro}
		Let $S_{0},P \in \mathit{PatPred}(\Omega,\Sigma)$ be
		constrained pattern predicates 
		in standard form with $\mathit{vars}(S_{0}) \cap
		\mathit{vars}(P)= \emptyset$.
		$\llbracket P \rrbracket$ is an invariant of 
		$(\mathcal{C}_{\mathcal{R},\mathit{State}},\rightarrow_{\mathcal{R}})$
		from $\llbracket S_{0} \rrbracket$ if: (i) 
		$S_{0} \subseteq P$, and (ii)
		$\mathcal{R}_{\mathit{stop}} \models_{[\,]}^{\forall} P
		\rightarrow^{\circledast} [P \sigma]$, where $\sigma$
		is a sort-preserving bijective renaming of
		variables such that
		$\mathit{vars}(P) \cap \mathit{vars}(P \sigma)= \emptyset$.
	\end{corollary}

\vspace{-1ex}

Corollary \ref{invariant-coro} can be extended to parametric
invariants (see \cite{maude-reach-log}). The treatment of 
co-invariants is similar and can also be found in \cite{maude-reach-log}.

\vspace{-2ex}

\section{A Sound Inference System} \label{sound-inf}

\vspace{-.5ex}

\newcommand{\logicrule}[2]{\displaystyle\frac{{\displaystyle #1}}{{\displaystyle #2}}}
\newcommand{\rrule}[2]{{#1} \longrightarrow^{\circledast} {#2}}
\newcommand{\ssequent}[3]{[{#1},\ {#2}] \ \mathrel{\vdash_{T}} \ {#3}}
\newcommand{\sequent}[4]{\ssequent{#1}{#2}{\rrule{#3}{#4}}}
\newcommand{\mlf}[2]{{#1} \mathrel{\mid} {#2}}
\newcommand{\myeqsymbol}[0]{\mathrel{=}_{E_\Omega \cup B_\Omega}}
\newcommand{\myeq}[2]{{#1} \mathrel{=}_{E_\Omega \cup B_\Omega} {#2}}
\newcommand{\termmodel}[0]{{\cal T}_{\Sigma/E \cup B}}
\newcommand{\mymodels}[1]{\termmodel \mathrel{\models} {#1}}
\newcommand{\match}[2]{\textsc{match}({#1},\ {#2})}
\newcommand{\unify}[2]{\textsc{unify}({#1},\ {#2})}
\newcommand{\cA}[0]{\ensuremath{\cal A}}
\newcommand{\cC}[0]{\ensuremath{\cal C}}
\newcommand{\vars}[1]{\mathit{vars}(#1)}
\newcommand{\nnmodels}[1]{\ensuremath{\models^{\forall,{#1}}_{T}}}
\newcommand{\sssequent}[3]{\R \mathrel{\vdash} [{#1},\ {#2}] \ {#3}}
\newcommand{\nmin}[0]{\ensuremath{n_\textit{min}}}
\newcommand\notmodels{\mathbin{\cancel{\models}}}

We present our inference system for all-path reachability for any
$\R$ satisfying topmost requirements (1)--(2),
with rules $R=\{ l_j \rightarrow
r_j \ \text{if} \ \phi_j \}_{j \in J}$ such that $ l_j , r_j \in
T_{\Omega}(X)$, $j \in J$.
\emph{Variables of rules in $R$ are always assumed disjoint from variables in
	reachability formulas}; this can be ensured by renaming.
The inference system has two proof rules. 
The $\textsc{Step}^{\forall} + \textsc{Subsumption}$ proof rule allows 
taking one step of (symbolic) rewriting along all paths according to the rules 
in $\R$. 
The \textsc{Axiom} proof rule allows the use of a trusted
reachability formula to summarize 
multiple rewrite steps, and thus to handle repetitive behavior.

These proof rules derive sequents of the form
$\ssequent{\cA}{\cC}{\rrule{\mlf{u}{\varphi}}{\bigvee_i \mlf{v_i}{\psi_i}}}$,
where $\cA$ and $\cC$
are finite sets of reachability formulas and
$T$ a pattern predicate defining a set of $T$-terminating ground states.
Formulas in $\cA$ are called {\em axioms} and those in $\cC$ are called {\em
	circularities}.  We furthermore assume that in all reachability formulas
$\rrule{\mlf{u}{\varphi}}{\bigvee_i \mlf{v_i}{\psi_i}}$ we have $\vars{\psi_i} \subseteq
\vars{v_i} \cup \vars{\mlf{u}{\varphi}}$ for each $i$. 
According to the implicit quantification of the semantic relation
$\models_{T}^{\forall}$ this means that any variable in $\psi_i$ is either universally
quantified and comes from the precondition $\mlf{u}{\varphi}$, or is existentially
quantified and comes from $v_i$ only.  This property is an invariant
preserved by the two inference rules.

Proofs always begin with a set $\cC$ of formulas that we want to
\emph{simultaneously} prove, so that the proof effort only succeeds
if \emph{all} formulas in $\cC$ are eventually proved.
$\cC$ contains the main properties we want to prove as well as
any auxiliary lemmas that may be needed to carry out the proof.
The initial set of goals we want to prove is $\ssequent{\emptyset}{\cC}{\cC}$, which is a
shorthand for the set of goals
$\{\ssequent{\emptyset}{\cC}{\rrule{\mlf{u}{\varphi}}{\bigvee_i \mlf{v_i}{\psi_i}}}\ \,
{\big |}\ \, (\rrule{\mlf{u}{\varphi}}{\bigvee_i \mlf{v_i}{\psi_i}}) \in \cC \}$.
Thus, we start \emph{without any axioms} $\cA$, but we shall be able
to use the formulas in $\cC$
as axioms in their own derivation \emph{after} taking at least on step
with the rewrite rules in $\R$. 

A very useful feature is that sequents
$\ssequent{\emptyset}{\cC}{\rrule{\mlf{u}{\varphi}}{\bigvee_i 
		\mlf{v_i}{\psi_i}}}$, whose formulas $\cC$ have been
\emph{postulated} (as the conjectures to be proved),
are transformed by $\textsc{Step}^{\forall} + \textsc{Subsumption}$
into sequents of the form
$\ssequent{\cC}{\emptyset}{\rrule{\mlf{u'}{\varphi'}}{\bigvee_i 
		\mlf{v'_i}{\psi'_i}}}$, where now the formulas in $\cC$
\emph{can be assumed valid}, and can be used in derivations with the
\textsc{Axiom} rule.

\vspace{1ex}

\noindent {\bf Verifying QLOCK's Mutual Exclusion}.  By Corollary
\ref{invariant-coro}, QLOCK's mutual exclusion can be verified
by: (i)  using pattern subsumption to check the trivial inclusion 
$\llbracket S_{0} \rrbracket \subseteq \llbracket P \rrbracket$, and (ii)
proving $\mathcal{R}_{\mathit{stop}} \models_{[\,]}^{\forall} P \sigma
\rightarrow^{\circledast} [P]$, where $\sigma$
is a sort-preserving bijective renaming of
variables such that $\mathit{vars}(P) \cap
\mathit{vars}(P \sigma)= \emptyset$.
But, since for QLOCK, $P$ is a disjunction, in our inference
system this means proving from $\mathcal{R}_{\mathit{stop}}$
that $[\emptyset,\mathcal{C}] \vdash_{[]} \mathcal{C}$,
where $\mathcal{C}$ are the conjectures: 
\[
\mathit{<n'\ |\ w'\ |\ i'\ |\ i'\ ;\ q'>} \ \! |\ \! \varphi' \rightarrow^{\circledast}
[ \mathit{<n\ |\ w\ |\ i\ |\ i\ ;\ q>} \ \! |\ \! \varphi \vee
\mathit{<n\ |\ w\ |\ \emptyset\ |\ q>} \ \! |\ \! \psi]
\]
\[
\mathit{<n'\ |\ w'\ |\ \emptyset\ |\ q'>} \ \! |\ \! \psi' \rightarrow^{\circledast}
[ \mathit{<n\ |\ w\ |\ i\ |\ i\ ;\ q>} \ \! |\ \! \varphi \vee
\mathit{<n\ |\ w\ |\ \emptyset\ |\ q>}\ \! |\ \! \psi].
\]
where $\varphi  \equiv \mathit{dupl}(\mathit{n\ w\ i})\neq \mathit{tt}$,
$\psi \equiv \mathit{dupl}(\mathit{n\ w})\neq \mathit{tt}$, and
$\varphi',\psi'$ are their obvious renamings.

Before explaining the $\textsc{Step}^{\forall} + \textsc{Subsumption}$
proof rule we introduce some notational conventions.
Assume $T$ is the pattern predicate
$T= \bigvee_j \mlf{t_j}{\chi_j}$, with $\vars{\chi_j} \subseteq
\vars{t_j}$, and let $R=\{ l_j \rightarrow
r_j \ \text{if} \ \phi_j \}_{j \in J}$, 
%
we then define:
{\small
	\[\match{u}{\{ v_i \}_{i \in I}} \subseteq \{ (i, \beta) \mid \beta \in [\vars{v_i}
	\setminus \vars{u} \rightarrow T_\Omega(X)] \text{ s.t. } \myeq{u}{v_i
		\beta}\}\]
}
a \emph{complete} set of (parameter-preserving)
$E_{\Omega} \cup B_{\Omega}$-matches of $u$ against the $v_i$, 
{\small
	\[\unify{\mlf{u}{\varphi'}}{R} \equiv \{ (j, \alpha) \mid \alpha \in
	\textit{Unif}_{E_{\Omega} \cup B_{\Omega}}(u,
	l_j) \ \text{and} \ (\varphi' \wedge \phi_j)\alpha \ \text{satisfiable in} \
	\termmodel \}\] 
}
a complete set of $E_{\Omega} \cup B_{\Omega}$-unifiers of a pattern $\mlf{u}{\varphi'}$ with the lefthand-sides
of the rules in $R$ with satisfiable associated
constraints.\footnote{In the current prototype implementation
	(see Section \ref{REFL-IMP}),  variant satisfiability
	makes constraint checking decidable.  Future versions
	will only assume $\vec{E}$ convergent modulo $B$
	for the equational part $E \cup B$
	of $\mathcal{R}$, so that
	satisfiability of such constraints will in general be undecidable.
	Unifiers whose associated constraints  cannot be proved unsatisfiable will then be
	included in $\unify{\mlf{u}{\varphi'}}{R}$ as a safe
	over-approximation.
	The same approach will apply to the, in general undecidable,
	checking of satisfiability/validity  for other constraints involved
	in the application of the $\textsc{Step}^{\forall} +
	\textsc{Subsumption}$ or $\textsc{Axiom}$
	rules below: they will be either over-approximated, or will become
	proof obligations to be discharged by an inductive
	theorem prover.}
Consider now the rule:

\paragraph{$\textsc{Step}^{\forall} + \textsc{Subsumption}$}
\[
\logicrule
{\bigwedge_{(j, \alpha) \in \unify{\mlf{u}{\varphi'}}{R}}
	\sequent{\cA \cup \cC}{\emptyset}
	{(\mlf{r_j}{\varphi' \wedge \phi_j}) \alpha}
	{\bigvee_i (\mlf{v_i}{\psi_i}) \alpha}}
{\sequent{\cA}{\cC}{\mlf{u}{\varphi}}{\bigvee_i \mlf{v_i}{\psi_i}}}
\]
where $\varphi' \equiv \varphi \wedge \bigwedge_{(i, \beta) \in \match{u}{\{ v_i \}}} \neg
(\psi_i \beta)$.
This inference rule allows us to take one step with the rules in $\R$.
Intuitively, $\mlf{u}{\varphi'}$ characterizes the states satisfying $\mlf{u}{\varphi}$ that
are not subsumed by any $\mlf{v_i}{\psi_i}$; that is, states in the
lefthand side
of the current goal that have not yet reached the righthand side.
Note that, according to Definition~\ref{all-paths-semantics},
$\rrule{\mlf{u}{\varphi}}{\bigvee_i \mlf{v_i}{\psi_i}}$ is semantically valid iff
$\rrule{\mlf{u}{\varphi'}}{\bigvee_i \mlf{v_i}{\psi_i}}$ is valid.
Thus, this inference rule only unifies $\mlf{u}{\varphi'}$ with the lefthand
sides of
rules in $R$.
We impose on this inference rule a side condition that $\bigvee_{j, \gamma \in
	\textit{Unif}_{E_{\Omega} \cup B_{\Omega}}(u, t_j)} (\varphi' \wedge \chi_j) \gamma$ is unsatisfiable in $\termmodel$,
where $T=\bigvee_j \mlf{t_j}{\chi_j}$ is the pattern predicate
characterizing the chosen
$T$-terminating states.
This condition ensures that any state in $\mlf{u}{\varphi'}$ has an
$\R$-successor.
Thus, a state in $\mlf{u}{\varphi'}$ reaches on all $T$-terminating paths a state in $\bigvee_i
\mlf{v_i}{\psi_i}$ if all its successors do so.
Each $\R$-successor is covered by one of $(\mlf{r_j}{\varphi' \wedge \phi_j}) \alpha$.
As an optimization, we check that $(\varphi' \wedge \phi_j) \alpha$ is satisfiable and we
drop the ones which are not.
Finally, we also assume that $\vars{(\mlf{u}{\varphi})\alpha} \cap
\vars{(\bigvee_i\mlf{v_i}{\psi_i})\alpha} = \vars{(\mlf{r_j}{\varphi' \wedge
		\phi_j})\alpha} \cap \vars{(\bigvee_i\mlf{v_i}{\psi_i})\alpha}$.
This parameter preservation condition ensures correct implicit quantification.
Note that formulas in $\cC$ are added to $\cA$, so that from now on they
can be used by \textsc{Axiom}.
By using $E_{\Omega} \cup B_{\Omega}$-unification, this inference rule
 performs narrowing of
$\mlf{u}{\varphi'}$ with rules $R$ \cite{narrowing-hosc}.

\paragraph{\textsc{Axiom}}
\[
\logicrule
{\bigwedge_j
	\sequent
	{\{ \rrule{\mlf{u'}{\varphi'}}{\bigvee_j \mlf{v'_j}{\psi'_j}} \} \cup \cA}
	{\emptyset}{\mlf{v'_j \alpha}{\varphi \wedge \psi'_j \alpha}}{\bigvee_i \mlf{v_i}{\psi_i}}}
{\sequent
	{\{ \rrule{\mlf{u'}{\varphi'}}{\bigvee_j \mlf{v'_j}{\psi'_j}} \} \cup \cA}
	{\emptyset}{\mlf{u}{\varphi}}{\bigvee_i \mlf{v_i}{\psi_i}}}
\]
if $\exists \alpha$ such that $\myeq{u}{u' \alpha}$ and $\mymodels{\varphi \Rightarrow
	\varphi' \alpha}$.
This inference rule allows us to use a trusted formula in $\cA$ to summarize
multiple transition steps.
This is similar to how several transition steps would apply to a ground term, except that for
ground terms we would check that $\varphi' \alpha$ is valid, whereas here we check that
the condition $\varphi$ implies $\varphi' \alpha$.
Since $\varphi$ is stronger than $\varphi' \alpha$, we add $\varphi$ to $(\mlf{v'_j}{\psi'_j})
\alpha$ (the result of using axiom  $\rrule{\mlf{u'}{\varphi'}}{\bigvee_j \mlf{v'_j}{\psi'_j}}$).
We assume that $\rrule{\mlf{u}{\varphi}}{\bigvee_i \mlf{v_i}{\psi_i}}$ and
$\rrule{\mlf{u'}{\varphi'}}{\bigvee_j \mlf{v'_j}{\psi'_j}}$ do not share variables, which
can always be guaranteed by renaming.
For correct implicit quantification, as in $\textsc{Step}^{\forall} +
\textsc{Subsumption}$, we assume for each $j$  the parameter preservation condition
$\vars{\mlf{u}{\varphi}} \! \cap\! \vars{\bigvee_i \mlf{v_i}{\psi_i}} =
\vars{\mlf{v'_j \alpha}{\varphi \wedge \psi'_j \alpha}} \! \cap\! \vars{\bigvee_i \mlf{v_i}{\psi_i}}$.
On a practical note, in order to be able to find the $\alpha$, our
implementation requires that $\vars{\varphi'} \subseteq \vars{u'}$, so that all
the variables in $\vars{\varphi'}$ are matched.

The soundness of $\textsc{Step}^{\forall} + \textsc{Subsumption}$ plus
\textsc{Axiom} is now the theorem:
\begin{theorem}\label{thm:soundness} (Soundness)
	Let $\R$ be a rewrite theory, and $\cC$ a finite set of reachability formulas.
	If $\R$ proves
	$\ssequent{\emptyset}{\cC}{\cC}$ then $\R \models_{T}^\forall \cC$.
\end{theorem}

Investigating  completeness of the logic is left as future work.

\vspace{-3ex}

\section{Prototype Implementation and Experiments} \label{REFL-IMP}

\vspace{-.5ex}

We have implemented the reachability logic proof system in Maude \cite{maude-book}.
Our prototype takes as input
(i) a  rewrite theory $\R=(\Sigma,E \cup B,R,\phi)$ and
(ii) a set of reachability formulas
$\mathcal{C} = \{ A_i \rightarrow^{\circledast} B_i \}_{i \in I}$ to be simultaneously proved.

To mechanize the two proof
rules we use a finitary $B$-unification algorithm as well as an
SMT solver to discharge $E\cup B$ constraints. 
%
%
For SMT solving we use 
variant satisfiability  \cite{var-sat-short,MLAVBS-WRLA16}, which
allows us to handle any rewrite theory $\mathcal{R}=(\Sigma,E\cup
B,R)$ satisfying topmost requirements (1)--(2) and
such that the equational
theory $(\Sigma,E\cup B)$ has a convergent
decomposition satisfying the finite variant property \cite{comon-delaune}
and protects a constructor subtheory which
we assume consists only of commutative and/or AC and/or identity
axioms $B_{\Omega}$.  
Thus, both validity and satisfiability of QF formulas in
the initial algebra $\mathcal{T}_{\Sigma/E \cup B}$ are decidable \cite{var-sat-short}.
Future implementations will support more general rewrite theories,
add other decision procedures, and use an
inductive theorem prover backend.

%
We have verified
properties for a suite of examples of rewrite theories
specifying distributed systems such as communication or mutual
exclusion protocols and real-time systems.
Table \ref{tab:table1} summarizes these experiments. 
For further
details plus runnable code see 
\url{http://maude.cs.illinois.edu/tools/rltool/}.

\begin{table}[h!]
	\centering
	\caption{Examples Verified in the Prototype Implementation}
	\label{tab:table1}
	{\tymax=250pt
		\begin{tabulary}{\linewidth}{LL}
			\toprule
			{Example \ \ } & {Description of the System/Property\ }\\
			\midrule
			Choice & Nondeterministically throws away elements from a multiset/eventually
			only one element left\\
			Comm. Protocol 1 & Simple communication protocol/received data is always a prefix of the data to be sent\\
			Comm. Protocol 2 & Fault-tolerant communication protocol/all data is eventually received in-order\\
			Dijkstra & Dijkstra's mutual exclusion alg./mutual exclusion\\
			Fixed-Size Token Ring & 2-Token ring mutual exclusion alg./mutual exclusion\\
			QLOCK & QLOCK mutual exclusion alg./mutual exclusion\\
			Readers/Writers & Readers-writers mutual exclusion alg./mutual exclusion\\	
			Lamport's Bakery & Unbounded Lamport's bakery/mutual exclusion\\
			Thermostat & Open system that dynamically responds to
			temperature/temperature remains in preset bounds\\
			\bottomrule
		\end{tabulary}
	}
\end{table}

\begin{figure}[h!]
$T_1 \equiv
\Bigg\{\begin{prooftree}
\infer0[\scriptsize sub$(P_1,\alpha)$]
{\parbox[c]{25em}{\centering
		$[\mathcal{C},\emptyset]\vdash_{[]} [n^3\ |\ w^3\ |\ \emptyset\ |\ q^3] \ \! |\ \! \mathit{dupl}(\mathit{n''\, w'\, p})\neq \mathit{tt}\ \wedge$\\ $\mathit{dupl}(\mathit{n^3 w^3})\neq \mathit{tt} \rightarrow^{\circledast}[P_1]\vee[P_2]$}
}
\end{prooftree}$\\

\vspace*{10pt}

$T_2 \equiv
\Bigg\{\begin{prooftree}
\infer0[\scriptsize sub$(P_2,\alpha)$ ]
{
	\parbox[c]{25em}{\centering $[\mathcal{C},\emptyset]\vdash_{[]} [n^3\ |\ w^3\ |\ i^3\ |\ i^3 ; q^3] \ \! |\ \! \mathit{dupl}(\mathit{n''\, w'\, p})\neq \mathit{tt}\ \wedge$\\ $\mathit{dupl}(\mathit{n^3 w^3 i^3})\neq \mathit{tt} \rightarrow^{\circledast}[P_1]\vee[P_2]$}
}
\end{prooftree}$

\vspace*{10pt}
\hspace*{15pt}
\begin{prooftree}
	\hypo{T_1}
	\hypo{T_2}
	\infer[separation=10em]2[\scriptsize axiom$(G_2,\alpha)$
	]{\cdots
		\parbox[c]{22em}{\centering
			$[\mathcal{C},\emptyset]\vdash_{[]}\mathit{<\!n''\ |\ w'\ p\ |\ \emptyset\ |\ q'\!>} \ \! |\ \! \mathit{dupl}(\mathit{n''\, w'\, p})\neq \mathit{tt}$ $\rightarrow^{\circledast}[P_1]\vee[P_2]$}
		\cdots
	}
	\infer1[\scriptsize step$(\textit{n2w},\theta)$]{
		\parbox[c]{22em}{\centering
			$[\emptyset,\mathcal{C}]\vdash_{[]}\mathit{<\!n'\ |\ w'\ |\ \emptyset\ |\ q'\!>} \ \! |\ \! \mathit{dupl}(\mathit{n'\ w'})\neq \mathit{tt}$\\
			$\rightarrow^{\circledast}$ $[P_1]\vee[P_2]$
		}
	}
\end{prooftree}
\caption{Partial Proof Tree for QLOCK}
\label{prooffig}
\end{figure}

\noindent
To illustrate how the tool works in practice,
Fig. \ref{prooffig} shows a partial derivation of a sequent.
Recall that for QLOCK we had to prove
$[\emptyset,\mathcal{C}] \vdash_{[]} \mathcal{C}$,
where $\mathcal{C}$ was two already-discussed reachability
formulas $G_i\equiv P'_i\rightarrow[P_1]\vee [P_2]$ for $i\in\{1,2\}$
with respective preconditions the renamed disjuncts $P'_{i}$, $1
\leq i \leq 2$
in
invariant $P_{1} \vee P_{2}$, and postcondition $[P_1]\vee [P_2]$,
where $P_1\equiv\mathit{<n\ |\ w\ |\ i\ |\ i\ ;\ q>}\ \! |\ \! \textit{dupl}(n\ w\ i)\neq tt$
and $P_2\equiv\mathit{<n\ |\ w\ |\ \emptyset\ |\ q>}\; \! |\; \! \textit{dupl}(n\ w)\neq tt$.
Now, consider
$[\emptyset,\mathcal{C}] \vdash_{[]} P'_2\rightarrow^\circledast [P]$.
In the proof fragement below, the initial
sequent must apply the step rule. The result of step$(n2w,\theta)$ is the goal resulting from
unifying the head of the sequent with the lefthand side of the rule $n2w$ using the unifier $\theta=\{n\mapsto n'' p, w\mapsto w', c\mapsto\emptyset, q\mapsto q'\}$.
The next inference axiom$(G_2,\alpha)$ applies axiom $G_2$ using the substitution $\alpha\supseteq\{n\mapsto n^3, w\mapsto w^3, i\mapsto i^3, q\mapsto q^3\}$.
Since $G_2$ has two constrained patterns in its succedent, we derive two new goals, represented by proof trees $T_1$ and $T_2$. In either case, we can immediately subsume by noting
that our reachability formula's antecedent is an instance of either $[P_1]$ or $[P_2]$ using substitution $\alpha$, thus terminating the proof.

\vspace*{10pt}

\section{Related  Work and Conclusions} \label{REL-CONCL}

\vspace{-.5ex}

This work
extends  reachability logic~\cite{DBLP:conf/rta/StefanescuCMMSR14,DBLP:conf/oopsla/StefanescuPYLR16} 
to a rewrite-theory-generic logic to reason about \emph{both} distributed system
designs and  programs.  This extension is non-trivial.  It requires:
 (i) relativizing
terminating sequences to a chosen subset $\llbracket T \rrbracket$
of terminating states; (ii) solving the ``invariant paradox,'' 
to reason about invariants and co-invariants and characterizing them by 
reachability formulas through a theory transformation; and (iii)
making it possible to achieve higher levels of automation by
systematically basing the state predicates on positive Boolean
combination of  patterns of the form
$u \mid \varphi$ with $u$ a constructor term. 

In contrast,
standard reachability logic
\cite{DBLP:conf/rta/StefanescuCMMSR14,DBLP:conf/oopsla/StefanescuPYLR16}
uses matching logic,
which assumes a first-order model ${\cal M}$ and its 
satisfaction relation ${\cal M} \models \varphi$ in its
reachability logic proof system.
As discusses in Section~\ref{CCP},
we choose $T_{\Sigma/E \cup B}$ as the model
and $\rightarrow_{\mathcal{R}}$ for transitions,
rather than some general
${\cal M}$
and systematically exploit the isomorphism
$T_{\Sigma/E \cup B}|_{\Omega} \cong T_{\Omega/E_{\Omega} \cup
	B_{\Omega}}$,
allowing us to use unification, matching, narrowing, and
satisfiability procedures
based on the typically much simpler initial algebra of constructors
$T_{\Omega/E_{\Omega} \cup B_{\Omega}}$.  This has the advantage that
we can explicitly give the complete details of our inference rules
(e.g. how $\textsc{Step}^{\forall} + \textsc{Subsumption}$ checks the
subsumption, or ensures that states have at least a successor),
instead of relying on a general satisfaction relation $\models$ on some
${\cal M}$.  The result is a simpler logic with only two
rules (versus eight in \cite{DBLP:conf/rta/StefanescuCMMSR14,DBLP:conf/oopsla/StefanescuPYLR16}). 

We agree with the work in \cite{DBLP:conf/birthday/LucanuRAN15}
on the common
goal of making reachability logic rewrite-theory-generic, but differ
on the methods used.  Main differences include: (1) 
\cite{DBLP:conf/birthday/LucanuRAN15} does not give an inference system
but a verification algorithm.  (2) the theories used in
\cite{DBLP:conf/birthday/LucanuRAN15} 
assume restrictions like those in
\cite{rw-SMT-JLAMP} for  ``rewriting modulo SMT,'' which limit the
class of equational theories.  (3) Matching is
used in \cite{DBLP:conf/birthday/LucanuRAN15} instead of
unification.  Thus,
unless a formula has been sufficiently
instantiated, no matching rule may exist, whereas unification with
some rule is always possible in our case.  
(4) No method for proving invariants is given in
\cite{DBLP:conf/birthday/LucanuRAN15}.

In conclusion, the goal of making reachability
logic a rewrite-theory-generic verification logic has been advanced. 
 Feasibility  has been
validated with a prototype and a suite of
examples.  Building a robust and highly effective
reachability logic tool for rewrite theories is a more
ambitious future goal. 

\vspace{2ex}

\noindent {\bf Acknowledgements}.  Partially supported by NSF
Grant CNS 14-09416.

\vspace{-2ex}

\bibliography{tex}

\end{document}